\documentclass[PP]{AEA}

\usepackage{mathptmx,comment, graphicx, subcaption, hyperref, chapterbib}

\usepackage[sectionbib]{natbib}
\usepackage{lipsum}

\draftSpacing{1.5}

\begin{document}

\title{Predicting Well-Being with Mobile Phone Data:\\Evidence from Four Countries}
\shortTitle{Measuring Poverty with Phone Data}
\author{Emily Aiken and Joshua E. Blumenstock and Sveta Milusheva and M. Merritt Smith\thanks{Aiken: U.C. San Diego (eaiken@ucsd.edu); Blumenstock: U.C. Berkeley (jblumenstock@berkeley.edu); Milusheva: World Bank (smilusheva@worldbank.org);
Smith: U.C. Berkeley (merrittsmith@berkeley.edu). We thank Oscar Barriga Cabanillas for assistance accessing data from Côte d'Ivoire, and James Foster for comments on an early draft. We gratefully acknowledge funding from the Global Data Facility Mobile Phone Data for Policy Program at the World Bank, which is funded by the Government of Spain.  The findings, interpretations, and conclusions expressed in this paper are entirely those of the authors. They do not necessarily represent the views of the World Bank and its affiliated organizations, or those of the Executive Directors of the World Bank or the governments they represent.
}}\pubMonth{Month}
\date{\today}
\pubYear{Year}
\pubVolume{Vol}
\pubIssue{Issue}
\maketitle

Accurate measurements of economic livelihoods are essential to empirical social science research, and to the design of effective policies. However, traditional survey-based approaches to measuring livelihoods are costly; as a result, many low- and middle-income countries (LMICs) lack timely and reliable data \citep{jerven2013poor}. 

Over the past decade, new digital data sources and advances in computation have enabled a new paradigm for estimating economic livelihoods. One growing line of research shows how mobile phone call detail records --- the metadata generated when phone calls, text messages, and other events occur on mobile phone networks --- can be used to predict the wealth and poverty of individual subscribers \citep{blumenstock_calling_2014,blumenstock2018estimating}. The intuition is that wealthier subscribers use their phones differently than poorer ones, and that given the right training data, machine learning algorithms can learn these differences. The predictions generated by these algorithms have in turn been used to generate granular poverty maps \citep{blumenstock2015predicting,steele_mapping_2017}, and determine eligibility for social protection programs \citep{aiken2022machine,aiken_program_2023,aiken_scalable_2025}.

In this paper, we use rich data from four countries --- Afghanistan, Côte d'Ivoire, Malawi, and Togo --- to systematically explore the potential and limitations of using mobile phone data to estimate economic livelihoods. In each country, we link survey-based measures of well-being from hundreds or thousands of individuals to comprehensive metadata capturing each of those individuals' patterns of phone use. We then conduct parallel, standardized analysis in each country to address several key questions that arise when implementing phone-based poverty prediction approaches: (i) Which measures of well-being can be most accurately measured using phone data? (ii) What types of mobile phone data are most useful for generating accurate predictions? (iii) How many individuals must be surveyed to train accurate prediction models? 

We document four main results.  First, some measures of poverty and vulnerability can be more accurately predicted than others: in particular, estimates of wealth and multi-dimensional poverty are more accurate than estimates of consumption, which in turn are more accurate than estimates of food security and mental health. Second, metadata from phone calls and text messages --- which contain rich information on location and mobility --- are more predictive than metadata from mobile internet usage, mobile money transactions, or airtime top-ups. Third, predictive accuracy increases with training sample size, with rapid improvements over the first 1,000-2,000 surveyed individuals, then slower (but continued) improvements through at least 4,500. Finally, predictive accuracy varies sharply across countries, driven in part by the heterogeneity of the underlying population. For instance, predictions are substantially more accurate when based on a nationally-representative population than on just a rural or urban sample.

\section{Data and Methods}
\label{sec:data}
\begin{table}[htbp]
\centering
\small
\caption{Summary of phone and survey data collection by country}
\renewcommand{\arraystretch}{1.2}
\label{tab:data_collection}
\begin{tabular}{@{}llllll@{}}
\hline
\textbf{Country} & \textbf{Dates of phone data} & \textbf{Date of Survey} & \textbf{\textit{N}} & \textbf{Sample Frame} & \textbf{Modality} \\
\textbf{Afghanistan} & Apr. - June 2018 & July - Oct. 2018 & 528 & Rural households in Balkh province & In-person \\
\textbf{Côte d'Ivoire} & Sep. 2023 & Sep. - Oct. 2023 & 773 & Nationally representative & Phone \\
\textbf{Malawi} & Sep. - Nov. 2024 & Oct. - Dec. 2024 & 5,469 & Five largest cities & In-person \\
\textbf{Togo} & Oct. - Dec. 2018 & Sep. 2018 - June 2019 & 4,345 & Nationally representative & In-person \\
\hline
\end{tabular}
\end{table}

We work with household survey data from four countries: Afghanistan, Côte d'Ivoire, Malawi, and Togo. For each of the households surveyed, we obtain the phone number of at least one household member and informed consent to match their survey responses to their mobile phone metadata, which we separately obtain from the mobile phone operator.\footnote{In Afghanistan, Malawi, and Togo phone numbers are from household heads if available and other household members if not; in Côte d'Ivoire phone numbers are from the survey respondent.} The metadata contain records of four transaction types: calls and text messages, airtime top-ups, mobile data usage, and mobile money transactions. Some data types are only available in a subset of countries; Table \ref{tab:data_collection} summarizes the key features of each dataset. Other important differences include sample size (ranging from 528 individuals in Afghanistan to 5,469 in Malawi), sample composition (rural-only in Afghanistan, urban-only in Malawi, and nationally-representative in Côte d'Ivoire and Togo), survey mode (over the phone in Côte d'Ivoire vs. in person elsewhere), and the duration of available phone metadata  (one month in Côte d'Ivoire vs. three months in the other three countries). Appendix A1 provides full details on all four survey and mobile phone datasets. 

Our results are based on machine learning experiments in which predictive models are trained to predict survey-based poverty and vulnerability outcomes using features derived from mobile phone metadata.\footnote{We use around 2,000 features derived from mobile phone metadata calculated with the Python library \href{cider}{https://global-policy-lab.github.io/cider-documentation/intro.html}} Across experiments, we train three models --- a LASSO regression, random forest and gradient boosting model --- tuning hyperparameters with five-fold cross-validation, select the best model (based on lowest RMSE) using a held-out validation set, and report the result of the best model on a separate, held-out test set. Appendix A5 provides full details on the model training and evaluation procedures. We report results from the four countries, based on experiments that vary the outcome variable, the size of the training dataset, and the mobile phone data features used.

\section{Results}
\label{sec:results}

\begin{figure}
\caption{Cross-Country Model Performance by Target and Data Available}
\label{fig:outcomes}
\includegraphics[width=\textwidth]{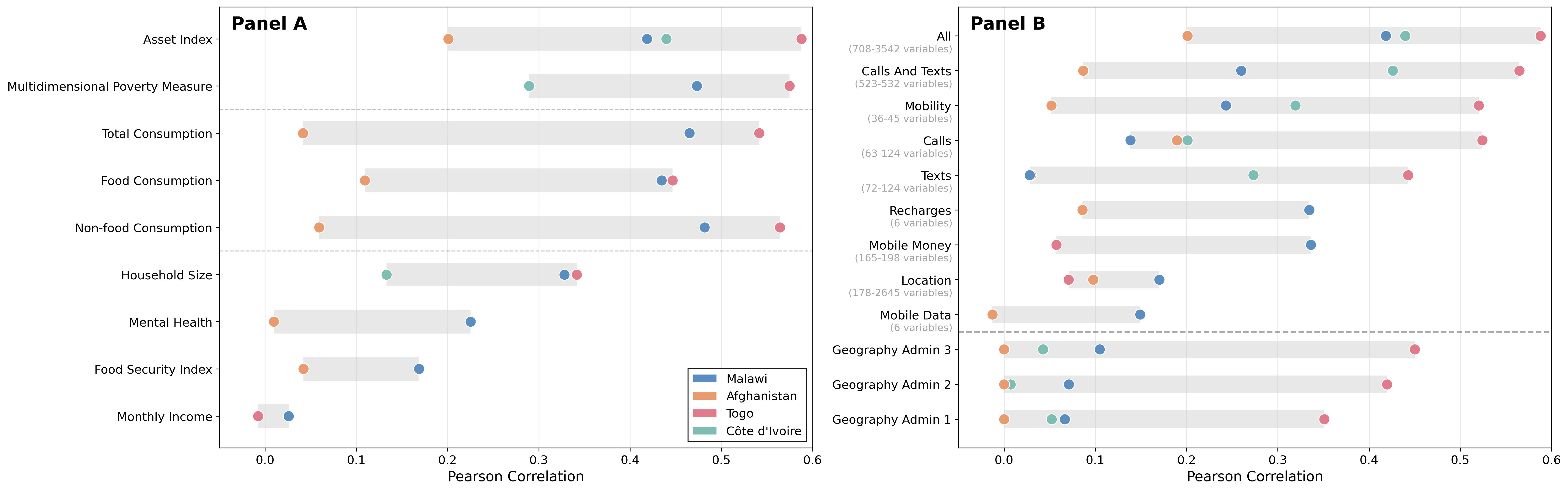}
\begin{figurenotes}
Panel A shows Pearson's $\rho$ between predicted and actual values across different poverty measures. Panel B shows performance using different mobile phone transaction types for predicting the asset index. Where data types are not available in certain countries, no results are shown in Panel B.  
\end{figurenotes}
\end{figure}

\begin{figure}
\caption{Varying Sample Size and Composition}
\label{fig:sample}
\includegraphics[width=\textwidth]{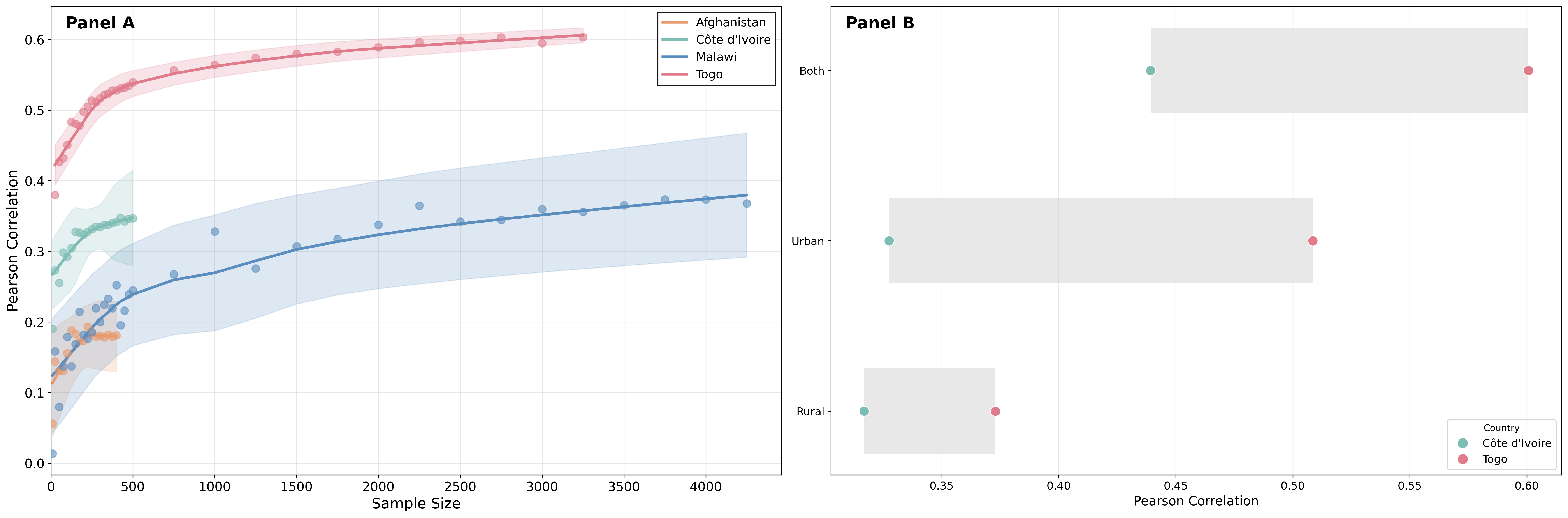}
\begin{figurenotes}
Panel A shows predictive accuracy (Pearson's $\rho$) as a function of the training dataset size, for predicting an asset index. Points are the mean holdout $\rho$ for the given sample size across bootstraps, and the shaded region is the 95\% confidence interval derived from 5 bootstrapped runs of the same sample size. Panel B shows predictive accuracy for the asset index in our two nationally representative samples (Côte d'Ivoire and Togo) when restricting to only rural and only urban individuals.  
\end{figurenotes}
\end{figure}

Figures \ref{fig:outcomes} and \ref{fig:sample} summarize the results of our experiments. In each figure, we present predictive accuracy (measured with Pearson's $\rho$ between predicted and observed values in the held-out test set) for a single country as a colored point; the range of predictive accuracy values across countries is highlighted in gray. For instance, in Figure \ref{fig:outcomes}A, the four dots in the top row indicate that the correlation between the true asset index value and the predicted asset index value (among the held-out test set) in Afghanistan is  0.20; in Malawi is 0.42; in Côte d’Ivoire is 0.44, and in Togo is 0.59.

One of the most striking patterns that emerges from the results in Figures \ref{fig:outcomes} and \ref{fig:sample} is that predictive accuracy is systematically higher in some countries, across all outcomes and data types. In particular, accuracy is highest in Togo and lowest in Afghanistan; with Malawi and Côte d’Ivoire somewhere in between. With only four countries, we do not base any conclusions on these  differences between countries, but later within-country analysis suggests that models perform better when the underlying sample is larger and more heterogeneous, which is consistent with Togo's diverse nationally representative sample and Afghanistan's narrow focus on rural households in a single province. 

In Figure \ref{fig:outcomes} Panel A, we assess the accuracy with which mobile phone data can predict a range of measures of poverty and vulnerability. Broadly, we observe that long-term poverty measures such as an asset index (Pearson $\rho$ = 0.20-0.59) and multidimensional poverty measure ($\rho$ = 0.29-0.57) are easier to predict than consumption ($\rho$ = 0.09-0.54), which is in turn easier to predict than income ($\rho$ = -0.01-0.03). Indices of vulnerability such as food security ($\rho$ = 0.04-0.17) and mental health ($\rho$ = 0.01-0.23) are also difficult to predict. We have reasonable accuracy for household size ($\rho$ = 0.13-0.34). 

In Figure \ref{fig:outcomes} Panel B, we explore which mobile phone data types are most predictive of poverty. We train our predictive models using only specific types of mobile phone data, focusing on predicting the asset index (Appendix Figure C3 replicates these results for consumption). In general, models that leverage only call data ($\rho$ = 0.11-0.52) or mobility data ($\rho$ = 0.05-0.52)  are nearly as predictive as models with all data ($\rho$ = 0.14-0.59). Mobile money ($\rho$ = 0.06-0.33), mobile data ($\rho$ = -0.01-0.14), recharges ($\rho$ = 0.09-0.32), and location data ($\rho$ = 0.07-0.14) are less predictive in isolation, although models that use all types of data consistently outperform models that use only specific types of data. To provide a naive baseline, the last three rows of Figure \ref{fig:outcomes} Panel B show the accuracy of machine learning models that use only location fixed effects from survey data. This is a useful baseline to compare against because it represents a simple approach to poverty estimation that could be taken in data-scarce settings: assign to a household the average poverty in the surrounding area, based on sample survey data. Across countries, the model leveraging all mobile phone data types substantially outperforms the naive location-based models. 

Our third experiment evaluates how much training data is necessary to train models to predict poverty from mobile phone data. Again using an asset index as the outcome, Figure \ref{fig:sample} Panel A demonstrates diminishing marginal returns to additional training data after around 1,000-2,000 individuals. However, ML model accuracy continues to improve past 4,000 individuals (our largest dataset size, in Malawi). Note that throughout our analysis, we have presented the results of the ``best'' machine learning model among the three we test --- LASSO regression, random forest, and gradient boosting -- where the best is the model selected on the held-out test set. To examine differences in training data needs across models, and to provide a more general cross-model accuracy comparison, Appendix Figure C4 shows a version of Figure \ref{fig:sample} Panel A for different model types. We find that the random forest and gradient boosting approaches have comparable accuracy, but that LASSO performs worse (except, in some countries, when very little data are available).

Figure \ref{fig:sample} Panel A also shows that when using a comparable sample size, Togo and Côte d’Ivoire significantly outperform Malawi and Afghanistan. Both of these countries use data from a nationally-representative sample, which is more heterogeneous than the data from Malawi (from five urban areas) and Afghanistan (from one rural province). In Panel B of Figure \ref{fig:sample}, we study explicitly the importance of sample heterogeneity, by testing performance of models on subsets of the training data. Specifically, we use our two nationally representative datasets (Togo and Côte d’Ivoire) to compare the ability of models to predict the asset index among just urban and just rural areas, relative to the full national sample. The national models ($\rho$ = 0.44-0.60) perform much better than the urban models ($\rho$ = 0.33-0.51) or the rural models ($\rho$ = 0.32-0.37). These results using mobile phone data are consistent with prior results showing that much of the predictive accuracy from satellite-based poverty maps comes from differentiating between rural and urban areas; when restricted to only urban or only rural areas, the accuracy of satellite-based poverty maps declines substantially \citep{aiken_fairness_2023}.


\section{Discussion}
\label{sec:discussion}
Mobile phone metadata create new opportunities for low-cost measurement of poverty and vulnerability at scale. Here, we provide evidence on the predictive performance of mobile phone metadata across countries, outcome measures, phone data types, dataset sizes, and sampling strategies. 

We conclude with a few policy implications that we expect will be useful to researchers and policymakers interested in using these methods. Most importantly, performance depends largely on the heterogeneity of the target population. In Togo and Côte d'Ivoire, performance increases by 20-70 percent when using the national sample rather than only urban or rural samples, and across countries, the models from countries with national samples (Togo and Côte d'Ivoire) perform much better than models from countries with more homogeneous samples (Malawi and Afghanistan). Second, using only standard call and text metadata produces models almost as accurate as those leveraging all mobile phone data types; in settings limited by computational complexity or data storage requirements, using only call and text data may be sufficient. Third, small training datasets (approximately 1,000 observations) are sufficient to train reasonably accurate models, though more data continues to improve model performance, particularly in homogeneous samples. Finally, mobile phone data appear best able to predict long-term poverty measures such as wealth and multi-dimensional poverty; predictions of more transient measures of vulnerability are less accurate.

There are several important limitations to our analysis. First, our work is based on data from four countries that differ in sample size, sample frame, timing, and survey instrument and modality --- and which are different socially, economically, and culturally. While we have standardized our analysis in an attempt to find generalizable patterns, some results may arise from the idiosyncrasies inherent to each country's dataset. Ideally, future research would incorporate more countries with standardized survey datasets, which could also shed light on other population characteristics that affect performance. Second, our focus is on the performance of models that predict the well-being of mobile phone subscribers; we do not consider the extent to which such predictions could be used to draw inferences about, or inform policies for, the full population --- including those without phones. In some settings, such as the targeting of mobile money transfers to phone owners \citep{aiken2022machine}, this may not create systematic exclusions; in others, such as the targeting of cash transfers to the rural poor \citep{aiken_program_2023}, it could systematically degrade program efficiency. Finally, our focus is on cross-sectional accuracy, since we have only one wave of survey data from each country. An exciting area for future research will explore the extent to which mobile phone data can detect changes in livelihoods over time, and provide more dynamic, real-time measures of well-being \citep{aiken2023moving}.

\vspace{1cm}
\bibliographystyle{aea}
\bibliography{targeting_bib}

\end{document}